\documentclass[superscriptaddress,prl,twocolumn,nofootinbib,showpacs,preprintnumbers,floatfix]{revtex4}

\usepackage{amsfonts}
\usepackage{mathrsfs}
\usepackage{amssymb}
\usepackage{amsmath}
\usepackage{graphicx,epsfig}

\def\bwt{\begin{widetext}}

\def\ewt{\end{widetext}}

\def\be{\begin{equation}}

\def\ee{\end{equation}}

\def\bea{\begin{eqnarray}}

\def\eea{\end{eqnarray}}

\def\bean{\begin{eqnarray*}}

\def\eean{\end{eqnarray*}}

\def\bary{\begin{array}}

\def\eary{\end{array}}

\def\bit{\begin{itemize}}

\def\eit{\end{itemize}}

\def\su5u1{SU(5) \times U(1)}

\def\fsu5u1{SU(5) \times U(1)'}

\def\so10{SO(10)}

\def\sq20{SO(10) \times SO(10)}

\usepackage{amssymb}

\begin{document}

\setlength{\parskip}{0cm}

\title{ Ultra High Jet Signals from Stringy No-Scale Supergravity }

\author{Tianjun Li}

\affiliation{George P. and Cynthia W. Mitchell Institute for Fundamental Physics and Astronomy, Texas A$\&$M University, College Station, TX
77843, USA }

\affiliation{Key Laboratory of Frontiers in Theoretical Physics,
      Institute of Theoretical Physics, Chinese Academy of Sciences,
Beijing 100190, P. R. China }

\author{James A. Maxin}

\affiliation{George P. and Cynthia W. Mitchell Institute for Fundamental Physics and Astronomy, Texas A$\&$M University, College Station, TX
77843, USA }

\author{Dimitri V. Nanopoulos}

\affiliation{George P. and Cynthia W. Mitchell Institute for Fundamental Physics and Astronomy,
 Texas A$\&$M University, College Station, TX 77843, USA }

\affiliation{Astroparticle Physics Group,
Houston Advanced Research Center (HARC),
Mitchell Campus, Woodlands, TX 77381, USA}

\affiliation{Academy of Athens, Division of Natural Sciences,
 28 Panepistimiou Avenue, Athens 10679, Greece }

\author{Joel W. Walker}

\affiliation{Department of Physics, Sam Houston State University,
Huntsville, TX 77341, USA }



\begin{abstract}

We present distinctive signatures of flipped F-Theory models with TeV-scale vector-like particles, a $\sqrt{s}$ = 7 TeV 1 $fb^{-1}$ test at LHC of a class of models well-motivated from string theory. The characteristic feature is a light stop and gluino, both sparticles lighter than all other squarks. This unique aspect of the supersymmetry spectrum generates an ultra-high multiplicity of hadronic jets. We find the optimal signal to background ratio is realized for 9 or more jets. Exclusion of the essential cuts presented here could leave the ultra-high jet signal severely attenuated and concealed.

\end{abstract}


\preprint{ACT-03-11, MIFPA-11-07}

\maketitle


{\bf Introduction~--}The Large Hadron Collider (LHC) at CERN has been accumulating 
data from ${\sqrt s}=7$ TeV proton-proton collisions since March 2010. It is expected 
to reach an integrated luminosity of $1~{\rm fb}^{-1}$ by the end of 2011, all 
in search of new physics beyond the
Standard Model (SM).  Supersymmetry (SUSY), which provides a natural 
solution to  the gauge hierarchy problem, is the most promising
 extension of the SM.  Data corresponding to a limited $35~{\rm pb}^{-1}$
has already established new constraints on the viable parameter
space~\cite{Khachatryan:2011tk, daCosta:2011hh} due to the unprecedented center of mass
collision energy.  The search strategy for SUSY signals in early LHC data has
been actively and eagerly studied by quite a few groups~\cite{Baer:2010tk},
 with particular focus
on the parameter space featuring a traditional mass relationship between squarks 
and the gluino, such as a gluino heavier than all squarks or a gluino lighter than all squarks.

A question of great interest is whether there exist SUSY models which are well 
motivated by a fundamental theory such as string theory, 
which can be tested in the initial LHC run, permitting 
a probe of the UV physics close to the Planck scale.
In this Letter we present 
such a model. It is well known that the supersymmetric flipped 
$SU(5)\times U(1)_X$ models can solve
the doublet-triplet splitting problem elegantly via the missing
partner mechanism~\cite{AEHN-0}. To realize the string scale gauge coupling
unification, two of us (TL and DVN) with Jiang proposed the testable 
flipped $SU(5)\times U(1)_X$ models with TeV-scale vector-like 
particles~\cite{Jiang:2006hf}, where such models can be 
realized in the ${\cal F}$-ree ${\cal F}$-ermionic string
constructions~\cite{Lopez:1992kg}
 and ${\cal F}$-theory model building~\cite{Jiang:2009zza}, dubbed $\cal{F}$-$SU(5)$.
In particular, we find the generic phenomenological consequences are quite
interesting~\cite{Jiang:2009zza, Li:2009fq}.

In Grand Unified Theories (GUTs) with
gravity mediated supersymmetry breaking, also known as the
mSUGRA model, 
the supersymmetry breaking soft terms can be parameterized
by four universal parameters: the gaugino mass $M_{1/2}$,
scalar mass $M_0$, trilinear soft term $A$, and
the ratio of Higgs vacuum expectation values $\tan \beta$ at low energy,
plus the sign of the Higgs bilinear mass term $\mu$.
The $\mu$ term and its bilinear 
soft term $B_{\mu}$ are determined
by the $Z$-boson mass $M_Z$ and $\tan \beta$ after
the electroweak (EW) symmetry breaking.
To address the cosmological flatness problem,
 No-Scale supergravity has been proposed~\cite{Cremmer:1983bf}. 
The simplest No-Scale boundary conditions $M_0=A=B_{\mu}=0$
 are automatically obtained from the simple K\"ahler potential given 
in~\cite{Li:2010ws}, while $M_{1/2} > 0$ is allowed and indeed required 
for SUSY breaking. However, historically this appealing reductionist perspective 
has been challenged by a basic inconsistency of the $M_0 = 0$ condition 
applied at a traditional GUT scale around $10^{16}$~GeV with precision 
phenomenology. 


Recently, with No-Scale boundary conditions at the $\cal{F}$-$SU(5)$ unification
scale, we described an extraordinarily constrained ``golden point''~\cite{Li:2010ws} 
and ``golden strip''~\cite{Li:2010mi} that satisfied all the latest 
experimental constraints and has an imminently observable proton 
decay rate~\cite{Li:2009fq}. In addition, exploiting a ``Super-No-Scale'' 
condition, we dynamically determined $M_{1/2}$ and $\tan\beta$.
Since $M_{1/2}$ is related to the modulus field of
the internal space in string models, we stabilized the modulus 
dynamically~\cite{Li:2010uu, Li:2011dw}. We discovered that the viable parameter 
space of $\cal{F}$-$SU(5)$ is quite narrow due to the stringent 
constraint from the $B_{\mu}=0$ condition. In the simplest 
No-Scale supergravity, 
all the SUSY breaking soft terms arise from a single 
parameter $M_{1/2}$, therefore, the spectra in the 
entire ``golden strip'' are very similar up to a small 
rescaling on $M_{1/2}$, with equivalent sparticle branching 
ratios, hence leaving invariant most of the ``internal'' 
physical properties, where this rescaling ability on $M_{1/2}$
 is not apparent in alternative SUSY models. 
For our analysis here, we use a vector-like particle mass of $M_V \sim 1000 $ GeV, which exists in the viable
parameter space; We emphasize that this is not an arbitrary choice~\cite{Li:2010mi}, though these vector-like
particles could be too heavy for observation in the early LHC run. Interestingly, the one-loop beta function 
for the $SU(3)_C$ gauge symmetry is zero due to the extra vector-like particle
contributions~\cite{Jiang:2006hf}, and consequently, the gluino is lighter than all the squarks except the light
stop. Furthermore, the gluino mass is about 600 GeV, so we can definitely test the
No-Scale $\cal{F}$-$SU(5)$ model at the early LHC run. 

In this analysis, we obey the strictest bottom-up phenomenological constraints 
and non-trivial consistency with equally strict top-down dynamical considerations,
and arrive ultimately at distinctive collider level signatures, including
a proposal for modest alterations to the canonical background selection cut 
strategy expected to yield significantly enhanced resolution of the characteristic ultra-high jet 
multiplicity $\cal{F}$-$SU(5)$ events. It is imperative we note that an ultra-high jet signal 
could be obscured by data selection cuts designed for low-multiplicity jet events. 
Therefore, we suggest that diligence against the cloaking of a prospective SUSY signal within the data collection already underway calls for attention to a broader range of anticipated spectra and cut selection criteria.

{\bf {Benchmark Point}~--}~To test our model, 
we present a
No-Scale ${\cal F}$-$SU(5)$
benchmark point in Table~\ref{tab:masses}.
The optimized signatures presented here offer an alluring testing 
vehicle for the stringy origin of $\cal{F}$-$SU(5)$.
This point is representative of the entire highly 
constrained ${\cal F}$-$SU(5)$ viable
parameter space. The SUSY breaking parameters 
for this point slightly differ from previous ${\cal F}$-$SU(5)$ 
studies~\cite{Li:2010ws,Li:2010mi,Li:2011dw} insomuch as more precise 
numerical calculations have been incorporated into our baseline algorithm.
The masses shift a few GeV from the spectra given in previous works, 
but we believe this to be a more accurate representation of 
an ${\cal F}$-$SU(5)$ spectrum, although the branching ratios
and decay modes of the spectrum in Table~\ref{tab:masses} in
this work and the spectra in~\cite{Li:2010ws,Li:2010mi,Li:2011dw} 
are identical, so the physical properties are consistent before and 
after code improvements. Thus, the signatures studied here will be 
common to the spectrum in this work as well as the spectra in previous papers.

\begin{figure*}[t]
        \centering
        \includegraphics[width=1.00\textwidth]{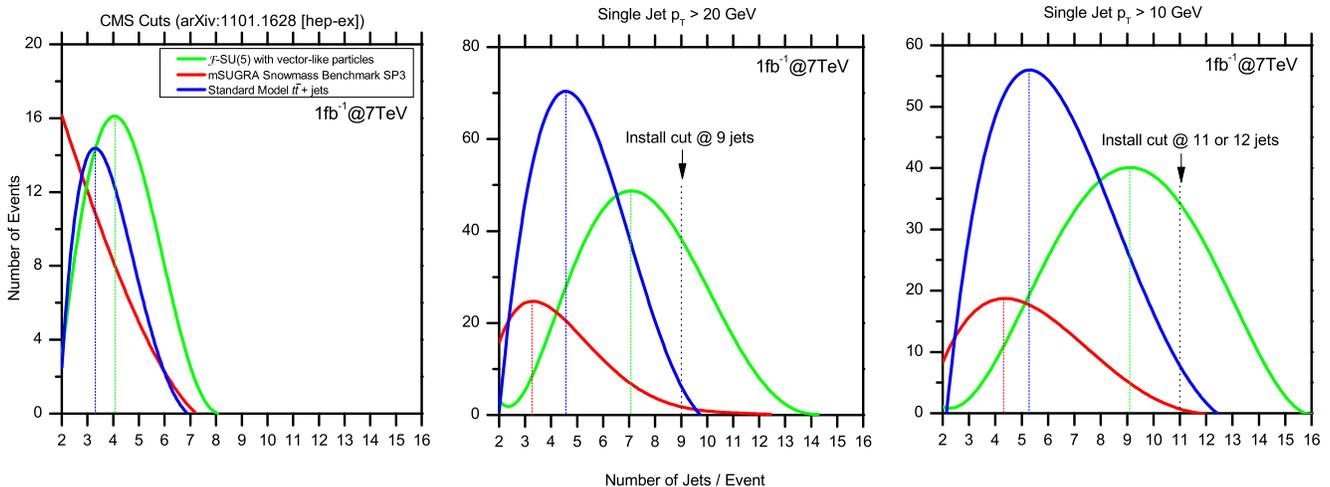}
        \caption{Distribution of events per number of jets. For clarity of the peaks, polynomials have been fitted over the histograms.}
\label{fig:jet_comp}
\end{figure*}

\begin{table}[ht]
  \small
	\centering
	\caption{Spectrum (in GeV) for $M_{1/2}$ = 410 GeV, $M_{V}$ = 1 TeV, $m_{t}$ = 174.2 GeV, tan$\beta$ = 19.5. Here, $\Omega_{\chi}$ = 0.11 and the lightest neutralino is 99.8\% bino.}
		\begin{tabular}{|c|c||c|c||c|c||c|c||c|c||c|c|} \hline		
    $\widetilde{\chi}_{1}^{0}$&$76$&$\widetilde{\chi}_{1}^{\pm}$&$165$&$\widetilde{e}_{R}$&$157$&$\widetilde{t}_{1}$&$423$&$\widetilde{u}_{R}$&$865$&$m_{h}$&$120.4$\\ \hline
    $\widetilde{\chi}_{2}^{0}$&$165$&$\widetilde{\chi}_{2}^{\pm}$&$756$&$\widetilde{e}_{L}$&$469$&$\widetilde{t}_{2}$&$821$&$\widetilde{u}_{L}$&$939$&$m_{A,H}$&$814$\\ \hline
    
    $\widetilde{\chi}_{3}^{0}$&$752$&$\widetilde{\nu}_{e/\mu}$&$462$&$\widetilde{\tau}_{1}$&$85$&$\widetilde{b}_{1}$&$761$&$\widetilde{d}_{R}$&$900$&$m_{H^{\pm}}$&$820$\\ \hline
    $\widetilde{\chi}_{4}^{0}$&$755$&$\widetilde{\nu}_{\tau}$&$452$&$\widetilde{\tau}_{2}$&$462$&$\widetilde{b}_{2}$&$864$&$\widetilde{d}_{L}$&$942$&$\widetilde{g}$&$561$\\ \hline
		\end{tabular}
		\label{tab:masses}
\end{table}

{\bf LHC Search~--}~For the initial phase of generation of the low order Feynman diagrams which may link the incoming beam to
the desired range of hard scattering intermediate states, we have used the program {\tt MadGraph 4.4}~\cite{Alwall:2007st}. These diagrams were subsequently fed into {\tt MadEvent}~\cite{Alwall:2007st} for appropriate kinematic scaling to yield batches of Monte Carlo simulated parton level scattering events. The cascaded fragmentation and hadronization of these events into
final state showers of photons, leptons, and mixed jets has been handled by {\tt PYTHIA}~\cite{Sjostrand:2006za}, with {\tt PGS4}~\cite{PGS4} simulating the physical detector environment. We implement MLM matching to preclude double counting of final states, and use the CTEQ6L1 parton distribution functions to generate the SM background. All 2-body SUSY processes are simulated. The b-jet tagging algorithm in {\tt PGS4} is adjusted to update the b-tagging efficiency to $\sim$60\%. We veto an event if any of the following conditions are met: $p_{T}$ $<$ 100 GeV for the two leading jets; $p_{T}$ $<$ 350 GeV for all jets; pseudorapidity $|\eta|$ $>$ 2 for the leading jet; missing energy $\mbox{E\!\!\!\!/}_{T}$ $<$ 150 GeV; isolated photon with $p_{T}$ $>$ 25 GeV; or isolated electron or muon with $p_{T}$ $>$ 10 GeV. Likewise, we discard any single jet with $|\eta|$ $>$ 3. These cuts are quite standard, but alone they are insufficient to reveal the ultra-high multiplicity jet event signature; We must also investigate the event cut on the number of jets and the $p_{T}$ cut on a single jet to preserve ultra-high jet events.

The detector simulations use the spectrum for the ${\cal F}$-$SU(5)$ point 
in Table~\ref{tab:masses}. The most significant asset of the spectrum for our analysis is the relationship between the stop, gluino, and other squarks. The distinctive mass pattern of $m_{\widetilde{t}_1} < m_{\widetilde{g}} < m_{\widetilde{q}}$ is the smoking gun signature and possibly a unique characteristic of only ${\cal F}$-$SU(5)$. To gain a comparison of the model studied here with more standardized SUSY models, we examine the ten ``Snowmass Points and Slopes'' (SPS) benchmark points~\cite{Allanach:2002nj} for suitable samples. We find that none of the ten SPS benchmarks support the $m_{\widetilde{t}_1} < m_{\widetilde{g}} < m_{\widetilde{q}}$ mass pattern. This critical element is indicative of how unique the ${\cal F}$-$SU(5)$ signal could be. Previous minimal supersymmetric SM studies focused on signals from a low-multiplicity of jets, whereas the aforementioned mass pattern is expected to show a very high-multiplicity of jets. For the SPS benchmarks, we only consider those spectra not light enough to have been excluded by the initial phase of LHC data, or those not too heavy for first year LHC production. A few points satisfy these criteria, though we select only one since we anticipate the corollary points to exhibit analogous characteristics. For our analysis here, we use the SPS SP3 benchmark.

Considering the large number of hadronic jets we are examining for our signatures, there is little intrusion from SM background processes after post-processing cuts. We examine the background processes studied in~\cite{Baer:2010tk,Altunkaynak:2010we} and our conclusion is that only the $t \overline{t} + jets$ possesses the requisite minimum cross-section and sufficient number of jets to intrude upon the ${\cal F}$-$SU(5)$ signatures. Processes with a higher multiplicity of top quarks can generate events with a large number of jets, however, the cross-sections are sufficiently suppressed to be negligible, bearing in mind the large number of ultra-high jet events which our model will generate. The same is true for those more complicated background processes involving combinations of top quarks, jets, and one or more vector bosons, where the production counts for 1 $fb^{-1}$ of luminosity are again sufficiently small. Furthermore, we neglect the QCD 2,3,4 jets, one or more vector bosons, and $b \overline{b}$ processes since none of these can sufficiently produce events with 9 or more jets after post-processing cuts have been applied.

\begin{figure}[ht]
        \centering
        \includegraphics[width=0.50\textwidth]{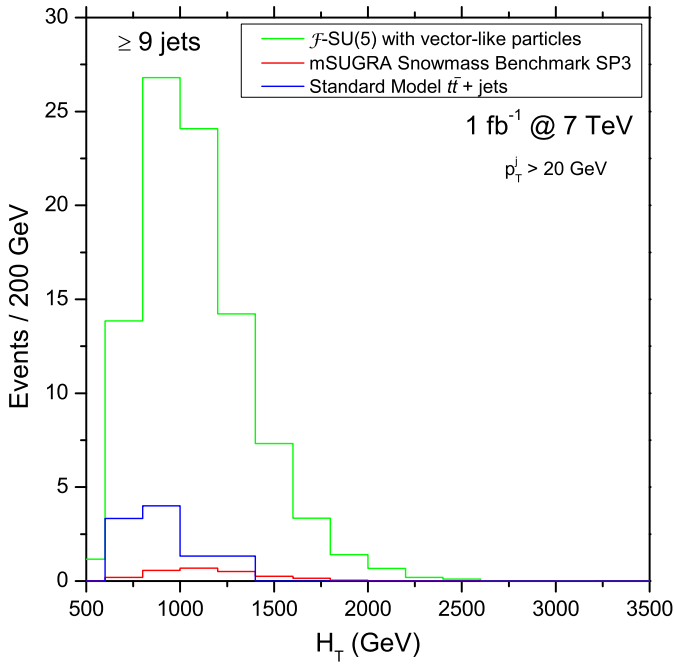}
        \caption{Counts for events with $\ge$ 9 jets.}
\label{fig:gte9jets}
\end{figure}

The ${\cal F}$-$SU(5)$ with vector-like particles mass pattern produces events with a high multiplicity of virtual stops, which concludes in events with a very large number of jets through the dominant chains $\widetilde{g} \rightarrow \widetilde{t}_{1} \overline{t} \rightarrow t \overline{t} \widetilde{\chi}_1^{0} \rightarrow W^{+}W^{-} b \overline{b} \widetilde{\chi}_1^{0}$ and $\widetilde{g} \rightarrow \widetilde{t}_{1} \overline{t} \rightarrow b \overline{t} \widetilde{\chi}_1^{+} \rightarrow W^{-} b \overline{b} \widetilde{\tau}_{1}^{+} \nu_{\tau} \rightarrow W^{-} b \overline{b} \tau^{+} \nu_{\tau} \widetilde{\chi}_1^{0}$, as well as the conjugate processes $\widetilde{g} \rightarrow \widetilde{\overline{t}}_{1} t \rightarrow t \overline{t} \widetilde{\chi}_1^{0}$ and $\widetilde{g} \rightarrow \widetilde{\overline{t}}_{1} t \rightarrow \overline{b} t \widetilde{\chi}_1^{-}$, where the $W$ bosons will produce mostly hadronic jets and some leptons. Additionally, the heavy squarks will produce gluinos by means of $\widetilde{q} \rightarrow q \widetilde{g}$. In Fig.~\ref{fig:jet_comp} we plot the number of jets per event versus the number of events for three distinct scenarios. We suppress the noise on the histogram contour to admit a more lucid distinction of the peaks in the number of jets, and fit polynomials over the data points and conceal the histograms. This allows us to gauge an appropriate selection cut for the number of jets to maximize our signal to background ratio, while assessing the impact of the selection cuts implemented by the CMS Collaboration in~\cite{Khachatryan:2011tk,PAS-SUS-09-001}. As depicted in Fig.~\ref{fig:jet_comp}, the first pane displays a comparison of the number of jets when employing the prior CMS cuts, while the remaining two panes present the results for the post-processing selection cuts defined in this paper, discriminating between two explicit cuts of the minimum $p_{T}$ for a single jet. Fig.~\ref{fig:jet_comp} demonstrates that the CMS cuts of~\cite{Khachatryan:2011tk,PAS-SUS-09-001} discard all the high-multiplicity jets, converting the events with at least 9 jets to events with few jets, thus, all information on these events with a large number of jets is lost. To retain the events with a high multiplicity of jets, we explore alternative cuts by shifting the minimum $p_{T}$ for a single jet lower to the two cases of 10 GeV and 20 GeV. A minimum jet $p_{T}$ of 20 GeV is secure from interfering with jet fragmentation, which typically occurs in the realm below 10 GeV, indicating that 10 GeV is certainly fringe. We see in Fig.~\ref{fig:jet_comp} that both the 10 GeV and 20 GeV jet $p_{T}$ cuts preserve the high number of jets, permitting an obvious choice for location of the cut on the minimum number of jets. We thus adopt a revised cut of single jet $p_{T} >$ 20 GeV and total number of jets greater than 9. To assess the discovery potential, we plot the number of events per 200 GeV versus $H_{T}$, where $H_{T} = \sum_{i=1}^{N_{jet}}E_{T}^{j_{i}}$. Fig.~\ref{fig:gte9jets} delineates the convincing separation between the ${\cal F}$-$SU(5)$ signal and the SM $t \overline{t} + jets$ and the SP3 point. The total number of events are summarized in Table~\ref{tab:counts}. We also include one measure of discovery threshold that compares the number of signal events S to the number of background events B, where we require $\frac{S}{\sqrt{B}} >$ 5. Notice that ${\cal F}$-$SU(5)$ comfortably surpasses this requirement.

\begin{table}[ht]
  \small
	\centering
	\caption{Total number of events for 1 $fb^{-1}$ and $\sqrt{s}$ = 7 TeV. Minimum $p_{T}$ for a single jet is $p_{T} >$ 20 GeV.}
		\begin{tabular}{|c|c|c|c|} \hline
		$$&${\cal F}$-$SU(5)$&$SP3$&$t\overline{t}+jets$\\ \hline\hline
    $Events$&$93.2$&$2.4$&$10$\\ \hline
    $\frac{S}{\sqrt{B}}$&$29.5$&$0.76$&$$\\ \hline
		\end{tabular}
		\label{tab:counts}
\end{table}

The spectrum of Table~\ref{tab:masses} exceeds the LEP constraints on the lightest neutralino $\widetilde{\chi}_{1}^{0}$ and lightest stau $\widetilde{\tau}_{1}$, and even more tantalizing, the close proximity of the stau mass beyond the LEP reach suggests imminent discovery at LHC. The stau presence can be reconstructed, for instance, from the dominant ${\cal F}$-$SU(5)$ process $\widetilde{g} \rightarrow \widetilde{t}_{1} \overline{t} \rightarrow b \overline{t} \widetilde{\chi}_1^{+} \rightarrow W^{-} b \overline{b} \widetilde{\tau}_{1}^{+} \nu_{\tau} \rightarrow W^{-} b \overline{b} \tau^{+} \nu_{\tau} \widetilde{\chi}_1^{0}$. The inference of the short-lived stau in the ${\cal F}$-$SU(5)$ SUSY breaking scenario from tau production assumes fruition of the expected much improved tau detection efficiency at LHC.

{\bf Conclusion~--}The LHC era has long been anticipated for the expected revelations of physics beyond the Standard Model, as the quest for experimental evidence and insight into the structure of the underlying theory at high energies is enticingly close at hand. Consequently, the field of prospective supersymmetry models has grown as fingerprints of these models at LHC are studied. Nevertheless, our exploration of recently published signatures for supersymmetry discovery reveals a common focus toward low-multiplicity jet events. However, we showed here that manipulation of LHC data skewed toward these low jet events could mask an authentic supersymmetry signal. We offer a clear and convincing ultra-high jet multiplicity signal for events with at least nine jets, unmistakable from the Standard Model or minimal supergravity. Notably, our optimized post-processing selection cuts outlined here are essential for discovery of supersymmetry if the ${\cal F}$-$SU(5)$ is indeed a physical model. Our revised cuts are not drastic, with the two chief adjustments being lowering the minimum $p_{T}$ for a single jet to 20 GeV, and raising the minimum number of jets in an event to nine. Recognition of such a signal of stringy origin at LHC could reveal not just the flipped nature of the high-energy theory, but also shed light on the geometry of the hidden compactified six-dimensional manifold in the string derived models. Thus, the stakes could not be higher or the revelations more profound.

{\bf Acknowledgments~--}~This research was supported in part 
by the DOE grant DE-FG03-95-Er-40917 (JM and DVN),
by the Natural Science Foundation of China 
under grant No. 10821504 (TL),
and by the Mitchell-Heep Chair in High Energy Physics (TL).


\end{document}